\documentstyle[12pt]{article}

\newcommand{\eq}{\begin{equation}}
\newcommand{\en}{\end{equation}}

\newcommand{\eqa}{\begin{eqnarray}}
\newcommand{\ena}{\end{eqnarray}}
\newcommand{\eqan}{\begin{eqnarray*}}
\newcommand{\enan}{\end{eqnarray*}}

\newcommand{\lbl}{\label}

\newcommand{\CMP}[1]{Comm. Math. Phys.\ {\bf #1}\ }

\newcommand{\IJMP}[1]{Int. Jou. Mod. Phys. \ {\bf #1}\ }

\newcommand{\MPL}[1]{Mod. Phys. Lett.\ {\bf #1}\ }

\newcommand{\NP}[1]{Nucl. Phys.\ {\bf #1}\ }

\newcommand{\PR}[1]{Phys. Rev\ {\bf #1}\ }

\newcommand{\RMP}[1]{Rev. Mod. Phys.\ {\bf #1}\ }
\newcommand{\ZP}[1]{Zeit. Phys.\ {\bf #1}\ }

\newcommand{\mat}[4]{\left( 
                     \begin{array}{cc}
                     {#1} & {#2} \\
                     {#3} & {#4} 
                     \end{array}
                     \right)
                    }

\newcommand{\matre}[9]{\left( 
                     \begin{array}{clcr}
                     {#1} & {#2} & {#3} \\
                     {#4} & {#5} & {#6} \\
                     {#7} & {#8} & {#9} 
                     \end{array}
                     \right)
                    }

\newcommand{\syst}[2]{\left\{
                      \begin{array}{c}
                     {#1} \\
                     {#2} 
                     \end{array}
                     \right.
                    }

\def\sqr#1#2{{\vcenter{\hrule height.#2pt
     \hbox{\vrule width.#2pt height#1pt \kern#1pt
        \vrule width.#2pt}
     \hrule height.#2pt}}}

\def\thinspace{\kern .16667em}

\newcommand{\ftt}[2]{{d^{#1}{#2}\over (2\pi)^{#1}} }

\def\part{\partial}

\def\sq2{\sqrt{2}}

\def\usq2{{1\over\sqrt{2}}}

\def\A{{\cal A}}
\def\B{{\cal B}}
\def\C{{\cal C}}
\def\D{{\cal D}}

\def\L{{\cal L}}

\def\U{{\cal U}}
\def\V{{\cal V}}
\def\Z{{\cal Z}}

\def\Dir{D\kern-1.7ex\Big{/}}
\def\Dsl{\partial\kern-1.5ex\Big{/}}

\def\ddm{\stackrel{\leftrightarrow}{\partial_-}}
\def\ddxm{\stackrel{\leftrightarrow}{\partial_{x^-}}}
\def\ddym{\stackrel{\leftrightarrow}{\partial_{y^-}}}
\def\ddu{\stackrel{\leftrightarrow}{\partial_1}}
\def\ddd{\stackrel{\leftrightarrow}{\partial_2}}

\def\reali{{\hbox{\s@ l\kern-.5ex R}}}
\def\naturali{{\hbox{\s@ l\kern-.5ex N}}}
\def\interi{{\mathchoice
 {\hbox{Z\kern-1.5mm Z}}
 {\hbox{Z\kern-1.5mm Z}}
 {\hbox{{Z\kern-1.2mm Z}}}
 {\hbox{{Z\kern-1.2mm Z}}}  }}

\def\unity{{\hbox{\s@ 1\kern-.8mm l}}}
\def\uno{{\hbox{ 1\kern-.8mm l}}}
\def\aa{\alpha}

\def\cc{\chi}
\def\cb{\bar\chi}
\def\dd{\delta}

\def\ee{\epsilon}

\def\ff{\phi}

\def\fd{\phi^{\dagger}}

\def\vf{\varphi}
\def\gg{\gamma}
\def\GG{\Gamma}

\def\ll{\lambda}
\def\LL{\Lambda}

\def\pp{\psi}

\def\pb{\bar\psi}

\def\rr{\rho}

\def\ss{\sigma}

\def\th{\theta}

\def\rmd{{r_-\over 2}}
\def \rad{\sqrt{1-{4m^2\over r^2}}} 
\def \radd{\sqrt{{4m^2\over r^2}-1}}

\def\bdpi{{b\over 2\pi}}

\begin{document}
\begin{titlepage}
\begin{flushright}
Bologna preprint DFUB-93-17\\
December 1993\\
hep-th/9401086
\end{flushright}
\vspace*{0.5cm}
\begin{center}
{\bf
\begin{Large}
{\bf 
A BILOCAL FIELD APPROACH TO THE LARGE-N EXPANSION OF 
    TWO-DIMENSIONAL (GAUGE) THEORIES
\\}
\end{Large}
}
\vspace*{1.5cm}
         {{\large Marco Cavicchi}\footnote{
         E-mail: cavicchi@bo.infn.it ~/ ~ cavicchi@nbivax.nbi.dk }
          }
         \\[.3cm]
          I.N.F.N. - {\it Sez. di Bologna and Dip. di Fisica,
         \\[.1cm]
          Universit\`a di Bologna, Via Irnerio 46, I-40126 Bologna,
          Italy}\\
\end{center}
\vspace*{0.7cm}
\begin{abstract}
{We consider a wide class of two-dimensional models as gauge theories, 
Gross-Neveu model, $O(N)$ and $CP^{N-1}$-like models using a formalism based on the 
introduction of bilocal fields that permits to perform easily the large-N 
expansion of this set of models in a unified and general way. We mainly
discuss the $SU(N)$ gauge field theory minimally coupled 
to fermionic plus bosonic matter in the fundamental representation,
and we obtain within the path integral
approach exact equations for the particle spectrum, also in presence of
renormalizable polynomial potentials.
Finally, we discuss the correspondence between this new approach and
the one previously used in the context of the $O(N)$ vector models.}
\end{abstract}
\vfill

\end{titlepage}
\setcounter{footnote}{0}


\setcounter{equation}{0}
\section{Introduction}

One of the most interesting and still not fully understood problem of
Quantum Field Theory concerns the calculation of the mass of relativistic 
bound states. This problem is even more complicated to solve in the case of 
$QCD$, the $SU(3)$ gauge theory of quarks and gluons, that describes the 
strong interaction physics where the basic constituents are confined and 
perturbation theory cannot be used. The expectation is that  the 
low energy spectrum consists of colorless mesons and baryons, but up to 
now the only available method seems to be based on numerical calculation 
using the lattice theory formulation of the theory.
 
The large-$N$ expansion technique proposed by 't Hooft~\cite{tH1} 
several years ago seems to be the most promising approach to obtain 
analytic results on the hadron spectrum.
It is well known that in the limit
in which the number $N$ of colours becomes very large the theory becomes 
much simpler, in the sense that only planar Feynman graphs survive.
Furthermore, when $N \rightarrow \infty$ the theory only contains colorless,
stable and noninteracting mesons with two-body decay and scattering amplitudes
proportional to ${1\over \sqrt{N}}$ and ${1\over N}$, respectively (there 
is also a way of describing baryons as solitons of the effective Lagrangian
in the large $N$ limit, but this will not interest us in this context).

Unfortunately, all efforts to try to solve four-dimensional large-$N$ 
$QCD$ have failed up to now.
The main problem is to find the semi-classical configuration,
namely the {\it master field}~\cite{Witten}, from which the action is 
dominated in the large-$N$ limit and whose fluctuations around the vacuum 
should give the particle spectrum of the theory. 

The main reason of this failure is due to the fact that no method 
has yet been found to solve matrix models for a space time dimension $D 
> 2$. It is well known that the large $N$ expansion can be explicitly 
performed in vector models as for instance the $O(N)$ vector model (see 
for instance ref.~\cite{PDV} and references therein), 
the two-dimensional $CP^{N-1}$ model~\cite{CPN}
and also in $QCD_2$~\cite{tH2},~\cite{CCG} with matter in 
the fundamental representation of the gauge group where the gluon field  
can be eliminated by using its equation of motion.


In particular this set of models are solved by means of two slightly 
different methods. The vector-like models are solved in the large $N$ 
expansion by introducing 
a local composite field and by explicitly integrating over the 
fundamental fields (See for instance Refs.\cite{PDV} and \cite{CPN}). 
The $QCD_2$-like models are characterized by the fact that the number of 
components of the non abelian gauge field goes to infinity in the large 
N expansion. The gauge field in two dimensions has no physical degrees 
of freedom and therefore can be eliminated by using its classical 
equation of motion. In this way one gets a non local Coulomb interaction 
that is quartic in terms of the vector-like matter fields. The theory 
can then be solved in the large N limit  by introducing a bilocal composite 
field as discussed in Refs.~\cite{GUT} and~\cite{MC}.

In this paper we consider a general two dimensional gauge theory with 
matter transforming as the fundamental representation of the gauge group 
(vector-like matter) and we solve it by introducing  bilocal colorless fields
in order to explicitly perform the large-$N$ expansion. The introduction 
of a bilocal field is essential because the gluon field has a number of 
components that goes to $\infty$ in the large $N$ expansion. We also 
show how our formalism reduces to the usual formalism of pure vector 
like models in the limit where the  gauge coupling constant  $g^2 
\rightarrow 0$.

The work is organized as follows. In section 2 we write  the Lagrangian
for a $SU(N)$ gauge theory minimally coupled to fermionic and bosonic 
matter in fundamental representation in the light-cone gauge.

In section 3  we consider the theory with only scalar fields
present, at first in the free massive case, then adding 
with a self-interaction potential $U(\fd\ff)$; we solve it in 
the large $N$ expansion and we show how to obtain from our solution 
the known results for the $O(N)$ model in the case in which the gauge coupling
$g$ is set to 0.
In the last subsection we constrain a (massless) interacting scalar
field to satisfy $\fd_i(x)\ff_i(x)=N/2f$, to find some similarities
with the $CP^{N-1}$ model.

In section 4  we consider the theory with fermionic fields, already studied in
~\cite{MC}, but adding current-current couplings; this kind of model is also
compared with the extension of the $CP^{N-1}$ model with quarks.

Finally in section 5 we study the full gauge theory, (fermionic plus
bosonic matter in fundamental representation of the gauge group) showing how to 
obtain the corresponding of the 't Hooft equation~\cite{tH2} for all particle 
sectors (b-b, b-f and f-f bound states)  
and examining the effect of adding any renormalizable and polynomial
self-interaction term.
We write exact integral equations, that reduce themselves, in the
case of zero gauge coupling, to the equations very similar to those previously
found for the spectrum of the massive Thirring model; in this case we find
the exact spectrum of states.
We also examine the case of nonzero Yukawa couplings between the 
fermionic and the bosonic $SU(N)$ fields.

\setcounter{equation}{0}
\section{Two-dimensional gauge theory with minimal coupling}

We start with the Lagrangian
\eq
\L= -{1\over 4} F^a_{\mu\nu}F^{a\mu\nu}
+\pb(i\Dir-m_0)\pp + (D_\mu\ff)^\dagger (D^\mu\ff)-m_0^2\fd\ff
\lbl{s-prime} 
\en
where $F_{\mu\nu}=\part_\mu A_\nu-\part_\nu A_\mu +ig_0 [A_\mu,A_\nu]$,
$\Dir_{A B}=\gg_\mu(\part_\mu \uno_{A B}+i g_0 A^a_\mu T^a_{A B})$.
All matter fields are in the fundamental representation of the gauge 
group $SU(N)$ and we have chosen  $T^a$ to be a Hermitian matrix.

Then we move to the light-cone coordinate system
\footnote{The conventions are the same as in~\cite{MC}}, 
and we interpret the $x^+$ coordinate as our `time' coordinate.
The Lagrangian is particularly simple in the light cone gauge 
$A^a_-=0$, because the nonabelian part of the field strength and his 
quadratic coupling with the boson field vanish:
\eq
\L_{l.c.} = {1\over 2} (\partial_-A^a_+)^2  
+\pb(i\Dsl-m_0)\pp + \fd(-\partial^2-m_0^2)\ff - g_0 A^a_+ J^a_- 
\en
with
\eq
J^a_- = \pb T^a \gamma_- \pp + i \fd T^a \ddm \ff
\lbl{ja}
\en
The equation of motion for the field $A^a_+$ does not contain any `time' 
derivative, so we can eliminate it to get an effective (non local) 
Action involving only the matter fields.
\eq
S_{eff} = \int d^2x d^2y\left\{
\left[ \pb(i\Dsl-m_0)\pp + \fd(-\partial^2-m_0^2)\ff\right]\dd_{xy} +
     {1\over 2} g_{0}^2 J^a_- {1\over \partial_-^2} J^a_- \right\}
\en

\setcounter{equation}{0}
\section{Bosonic theory}

\subsection{The 't Hooft equation for the bosonic sector}

We now want to study the theory in the case in which only scalar fields 
are present; the case with only fermions in the theory has already been 
treated~\cite{GUT,MC} with the same methods.
The model without field self-interaction, sometime also called "Scalar QCD", 
has been extensively studied~\cite{oenne} using the same techniques 
developed by 't Hooft; we now want to derive the same results by our
new approach.
We rewrite the interaction term in the following form:
\eq
S_{int} = {1\over 2}g_{0}^2 \int d^2x d^2y
\left(\fd(x)T^a\ddxm \ff(x)\right) G(x-y) 
\left(\fd(y) T^a \ddym \ff(y)\right)
\lbl{lint}
\en
with the Green's function $G$ defined by
\eq
-\partial_-^2 G(x-y) = \delta^{(2)}(x-y)
\en
so that in momentum space $G(k) = 1 / k_-^2 $. 
In the following, for the sake of simplicity in the notation, we will use 
$\partial_1$ and $\partial_2$ instead of $\partial_{x^-}$ 
and $\partial_{y^-}$, respectively.

By using the relation, that is valid only for fields transforming 
according to the fundamental representation~\cite{MC} 
\eq
\sum_a T^a_{A B} T^a_{C D}=\dd_{B C} \dd_{A D}
 -{1\over N}\dd_{A B} \dd_{C D}
\lbl{tata}
\en
we see that it is convenient to introduce as in~\cite{MC} a bilocal 
colour singlet field 
\eq
\ss^{xy} \equiv \sum_A \fd_A(y)\ff_A(x)
\en
The O(1/N) term in eq. (\ref{tata}) is completely irrelevant to our purposes, 
and we will neglect it in the following. 

With the help of (\ref{tata}) we can rewrite (\ref{lint}):
\eq
\L_{int} = {1\over 2} g_0^2 G_{xy} \ss^{xy}\, i\ddu i\ddd \ss^{yx} 
\en
The jacobian of the change of variable from $\fd \ff$ to $\ss $
can easily be computed in the large-N limit (see~\cite{DAS}, for 
example), and gives a contribution to the action that can be written 
formally as follows
\eq
\dd S[\ss] = -i N Tr \log\ss 
\en
Now, rescaling $\ss\rightarrow N\ss$ and sending $N\to \infty$ keeping
$g_{0}^2N=g^2$ fixed, we see that the action is dominated by a very simple 
saddle point.
The action in the large-N limit is then given by:
\eq
S/N = -i Tr \log\ss +\int d^2x d^2y \left[ 
-\dd_{xy}(\partial_x^2+m_0^2)\ss^{xy} + {1\over 2} g^2 G_{xy}\ss^{xy}
 i\ddu i\ddd \ss^{yx} \right]
\lbl{s-coord}
\en 
in configuration space and by
\eqa
 &&S/N = -i Tr \log\ss +\int \ftt{2}{p}\ftt{2}{q} 
(p^2-m_0^2)\ss(p,q)(2\pi)^2\dd^{(2)}(p+q)  
\nonumber\\
&&+{g^2\over 2} \int \ftt{2}{p_1}\ftt{2}{p_2}\ftt{2}{k}
{(2p_1+k)_-(2p_2-k)_-\over k_-^2}\ss(p_1,p_2)\ss(k-p_2,-k-p_1)
\nonumber\\
\lbl{s-impulse}
\ena 
in momentum space.\footnote{
We must give a suitable definition of $-\partial_-^{-2}$ or, in
momentum space, of $1/k_-^2$. We can easily see that the equation
$$
-\partial_-^2 G(x-y) = \dd^{(2)}(x-y)
$$
has the general solution
$$
G(x) = \dd(x^+)\left( -{1\over 2}|x^-| - Bx^- + A\right)
$$
The terms $A$ and $Bx^-$ are the zero-modes of the operator $\partial_-^2$.
As it has been pointed out by Einhorn~\cite{Einhorn},
the $B$ term corresponds to a background colored electric field, and without 
loss of generality it can be taken as zero; furthermore, the constant $A$ 
can be gauged away, so that we can conclude that it contains no physics.
In his work, 't Hooft~\cite{tH2} regularized his integrals by cutting off
all momenta $|p_-|<\ll$ where $\ll$ was a constant to send to zero at the end
of all calculations. It can be demonstrated that this procedure corresponds 
to the choice for the constant $A$ to be
equal to $1/\pi\ll$ so that in momentum space
$$
{1\over k_-^2}  = 
 {1\over 2} \left[{1\over (k_-+i\ee)^2}+{1\over (k_--i\ee)^2} \right] 
+ {1\over \pi\ll} (2\pi)\dd(k_-) 
\equiv  P{1\over k_-^2} + {2\over \ll} \dd(k_-) 
$$
In the following, for those "historical" reasons, we will keep this 
regularization prescription.}

The master field $\ss_0$  is the bilocal field configuration that 
dominates the functional integral for large $N$. It satisfies the 
saddle point of the action $ {\dd S\over \dd\ss }\biggr\vert_{\ss=\ss_0} = 0$;
As explained in Ref.~\cite{MC} translational invariance imposes to the 
master field to be local i.e. 
\eq
\ss_0(p,q) = \ss_0(p)(2\pi)^2\dd^{(2)}(p+q)
\en
A simple algebra permits to rewrite the master field equation as follows
\eq
\left(p^2-m_0^2-g^2\int\ftt{2}{k}{(2p_-+k_-)^2\over 
k_-^2}\ss_0(p+k)\right)\ss_0(p) = i
\lbl{mfeq}
\en
Defining
\eq
\dd m^2 \equiv g^2\int \ftt{2}{k}\ss_0(k)
\en
and using the following ansatz
\eq
\ss_0(p) = i(p^2-m^2+\GG(p)+i\ee)^{-1} 
\lbl{ss0}
\en
with $m^2 = m_0^2 + \dd m^2 $, we get  the equation for $\GG(p)$
\eq
\GG(p) = -g^2\int \ftt{2}{k} {(2p_-+k_-)^2-k_-^2 \over k_-^2}
      {i\over (p+k)^2-m^2+\GG(p+k)+i\ee}
\lbl{gamma-eq}
\en
that has the solution~\cite{tH2}
\eq
\GG(p) = {g^2\over \pi} - {g^2\over\pi}{|p_-|\over\lambda}
\lbl{gammap}
\en

In the case of scalar fields, unlike the fermionic case, 
we needed to renormalize the mass (the self-mass is affected by the 
ultraviolet divergence of the boson loop). 

The particle spectrum of the theory is obtaining by studying small 
fluctuation around the master field.  Putting $\ss = 
\ss_0 + {1\over \sqrt{N}}\dd\ss $, we find the effective action for the 
fluctuations by keeping only the quadratic part in $\dd\ss$,
from which we get the equation for the fluctuations:
\eq
\dd\ss(p,q)=ig^2\ss_0(p)\ss_0(-q)\int\ftt{2}{k}
{(2p+k)_-(2q-k)_-\over k_-^2}\dd\ss(k+p,-k+q)
\en
The bound state equation can now be obtained by simply following the 
same procedure as in Refs.~\cite{tH2},~\cite{MC}. One defines 
 $\dd\tilde\ss(r,s)=\dd\ss(p+q,{(p-q)/ 2}) $ and after the integration of 
both sides of the previous equation over $ ds_+/2\pi $, the previous
equation becomes (let us choose $r_->0$):
\eqa
&&\left[ {M^2\over 2|s_-+{r_-\over 2}|} + 
      {M^2\over 2|s_--{r_-\over 2}|} +
      {g^2\over\pi\lambda}-r_+\right]\vf(s_-) 
\nonumber\\
&&= {g^2\over 2\pi}\int_{s_--{r_-\over 2}} ^{s_-+{r_-\over 2}}
{dk_-\over k_-^2} {(2s_-+r_-+k_-)(-2s_-+r_--k_-)\over 
2|s_-+{r_-\over 2}| 2|s_--{r_-\over 2}| }\, \vf(s_-+k_-)
\nonumber\\
\ena
The $\vf$ field is simply defined as
\eq
\vf(s_-)=\int \ftt{}{s_+}\, \dd\tilde\ss(r,s)
\lbl{phij}
\en
(the $r$-dependence has been left implicit) 
and by definition $r$ is the total momentum of the two-parton system.

The $\ll$-dependent part cancels applying our regularization
\eq 
\int {dk_-\over k_-^2}\, \vf(s_-+k_-) = {2\over\ll}\vf(s_-) 
+ P \int {dk_-\over k_-^2}\, \vf(s_-+k_-)
\en
Rescaling and shifting the variables, we then easily get the 't Hooft 
equation for the bosonic sector ($r^2=2r_-r_+$):
\eq
 r^2 \vf(x) = \left[{M^2\over x} + {M^2\over (1-x)} \right] \vf(x)
 - {g^2\over \pi} P \int_0^1 {dy\over (y-x)^2}
{(x+y)(2-x-y)\over 2x\, 2(1-x)}\, \vf(y)
\lbl{thb}
\en

\subsection{Self-interacting scalar field}

In this subsection, we want to analyze the effects of adding a potential 
$ V(\fd\ff)$. In $D=2$ dimensions we can choose an arbitrary potential:
\eq
V(\fd\ff) = \sum_{n=2}^M {c_k\over k!}(\fd\ff)^k
\lbl{pot}
\en
without losing the renormalizability of the theory.

After the introduction of the bilocal field a potential term adds to the 
action a local term of the form 
\eq
\dd S = -\sum_{n=2}^M {c_k\over k!}\int d^2x d^2y \dd_{xy} (\ss_{xy})^k
\lbl{s-self}
\en
After the rescaling $\ss\rightarrow N\ss $ we see that in the large $N$ 
expansion we have to keep 
constant the quantities $ \bar c_{k} = c_k N^{k-1} $.
For simplicity we will only study a ${c_2 \over 2}(\fd\ff)^2$ potential,
and let it be $v=\bar c_2 =c_2 N$.\\
The new equation for the saddle-point is quite simple:
\eq
\left(p^2-m_0^2-g^2\int\ftt{2}{k}{(2p_-+k_-)^2\over 
k_-^2}\ss_0(p+k) - v \int\ftt{2}{k} \ss_0(k) \right)\ss_0(p) = i
\lbl{mfeq2}
\en
and it can be seen that the effect of the potential term on (\ref{mfeq})
is just to change the definition of the mass shift $\dd m^2$:
\eq
\dd m^2 \equiv (g^2+v)\int \ftt{2}{k}\ss_0(k)
\lbl{ren}
\en

We can then easily compute, as in the previous section, the equation for 
the fluctuations around the master field, 
getting the equation previously written by Ambj\o rn~\cite{Ambjorn}
($M^2 \equiv m^2- g^2/\pi$)

\eqa
 r^2 \vf(x) = \left[{M^2\over x} + {M^2\over (1-x)} \right] \vf(x)
  &-& {g^2\over \pi} P \int_0^1 {dy\over (y-x)^2}
{(x+y)(2-x-y)\over 2x\, 2(1-x)}\, \vf(y)
\nonumber\\
 &+&  {v\over 4\pi}{1\over x(1-x)} P \int_0^1 dy \, \vf(y)
\lbl{coupled}
\ena
that is equal to eq. (\ref{thb})  with the addition of the last term 
coming from the quartic potential.

If $g=0$ the gluon field, that is a matrix, decouples from the matter 
and we are left only with vector-like fields and  it is well known that 
one needs to introduce only a 
local composite field (see for instance Ref.~\cite{PDV}) and not a 
bilocal one. Therefore if we set $g=0$ the two methods must agree and 
this is what we are going to check in the following. 

Let us start from the master field equation in eq. (\ref{mfeq}) for the case 
$g=0$ whose solution is given by the expression in eq. (\ref{ss0}) with 
$\Gamma (p) =0$. Integrating both terms of eq. (\ref{ss0}) over $p$ and 
defining
\eq
-i \sigma_0 \equiv \delta m^2 = v \int \frac{d^2k}{(2\pi)^2} \sigma_0 
(k)
\lbl{gap1}
\en
we get, after a Wick rotation, the following equation
\eq
\frac{\sigma_0}{iv} = \int \frac{d^2 p}{(2\pi)^2} \frac{1}{-p^2 + m^2 -i 
\sigma_0}
\lbl{gap2}
\en
that is identical to equation (2.25) of Ref.~\cite{PDV} with 
$4f=v,K_f=K_n=0$. The r.h.s of eq. (\ref{gap2}) is divergent and must be 
regularized. The renormalized equation is obtained by extracting the 
divergent piece from $\sigma_0$ ( this divergent is due to the fact that 
$\sigma_0$ is the vacuum expectation value of a composite local field 
where the two constituent fields are taken at the same point) and having 
it to be cancelled by the divergence appearing in the r.h.s. of eq. 
(\ref{gap2}). After this procedure one obtains a renormalized gap 
equation ( See eq. (3.15) of Ref.~\cite{PDV}). In conclusion we have 
shown that, if $g=0$ the master field equation reduces to the gap 
equation of the vector models.

Let us consider now the fluctuation equation for $g=0$  
\eq
\left[ r^2-{m^2\over x(1-x)}\right] \vf(x) = {v\over 4\pi}{1\over 
x(1-x)} P \int_0^1 dy \, \vf(y)
\en
whose solution is given by
\eq
\vf (x) = {v\over 4\pi} {1\over  r^2\, x(1-x)-m^2} P\int_0^1 dy \, \vf(y)
\lbl{phixbb}
\en
By integrating over $x$ we get the consistency condition
\eq
1 = {v\over 4\pi} P\int_0^1 {dx\over r^2 \, x(1-x)-m^2}
\equiv {v\over 4\pi}F(m,r^2)
\lbl{con}
\en
that gives three cases, depending on the region in which we are looking for:
\\
i) $r^2>4m^2$: (\ref{con}) becomes

\eq
1 = {v\over 4\pi } {4\over r^2\rad}\,{1\over 2}\log {1+\rad \over 1-\rad}
\lbl{rgtm}
\en
or, parametrizing $r=2m\cosh\gg$, $\gg>0$ :

\eq
{\sinh \,2\gg \over 2\gg} = {v\over 4\pi m^2}
\en
that has an unique solution iff  $v > 4\pi m^2$, where we have a
resonance of mass $\mu^2 = \bar r^2$ with $\bar r^2 $ that satisfies 
(\ref{rgtm}).
\\
ii) $0<r^2<4m^2$: (\ref{con}) becomes

\eq
1 = -{v\over 4\pi } {4\over r^2\radd}{\rm Atg}{1\over\radd}
\lbl{rltm}
\en
that is, parametrizing  $r=2m\sin\th$, $0 < \th < \pi/2$ :
\eq 
{\sin\,2\th \over 2\th} = -{v\over 4\pi m^2}
\en
This condition can be satisfied only if $ -1< v/4\pi m^2 < 0 $ (see Fig.1),
and we have a bound state determined by (\ref{rltm}).

iii) $r^2<0$.
Obviously, if we find tachyonic solutions of (\ref{con}) the theory
is not defined. Writing $r^2=-4m^2\nu^2$, $\nu>0$ we get
\eq
\nu = -{v\over 4\pi m^2}{1\over \sqrt{1+\nu^2}}\log(\nu+\sqrt{1+\nu^2})
\en
that admits solution iff $ v/4\pi m^2 < -1 $ (see Fig.2),
so that this range of values for the coupling constant 
must be avoided\footnote{
This result agrees with the spectrum obtained using 
semiclassical methods by Abbott  ~\cite{Abbott}, who pointed out that the 
condition $ v/4\pi m^2 > -1 $ was necessary in order for a vacuum of the 
theory to exist (see also~\cite{Schnitzer} for a more detailed 
discussion).}.   

We can find a correspondence between the approach proposed here and
the methods that people used in the past.
In the usual procedure for the large $N$ expansion in the context of the 
$O(N)$ vector model one gets a quadratic term for the fluctuations 
around the solution of the gap equation that can be written as  
(eq. (2.29) of Ref.~\cite{PDV} )

\eq
S^{(2)}_{eff} = \frac{1}{2} \int \ftt{2}{k}\aa(k) \GG(k^2) \aa(-k)
\lbl{p-s2}
\en

$\GG(k^2)$ (not to be confused with our $\GG(p_-)$!) is defined by 
eq. (2.30) of Ref.~\cite{PDV}.
The equation of motion for the Fourier transform of the fluctuation field 
$\aa(x)$ gives $\GG(k^2)=0$ (with $\GG(k)$, in the $D=2$ case, that is 
given by eq. (3.30) of Ref.~\cite{PDV}).
This condition is just equal to eq. (\ref{rgtm}), Wick rotated to Euclidean 
space and with $v=4f$.

Another more direct way to find the connection between our approach and 
the one in terms of a local composite is the following. 
If we consider the equation of motion for the field $\dd\ss$
obtained from the action (\ref{s-impulse}) plus the self-interaction term,
the Fourier transform of (\ref{s-self}), setting $g=0$ we have

\eq
\dd\ss(p,q) = -iv\ss_0(p)\ss_0(-q)\int\ftt{2}{k}\dd\ss(k+p,-k+q)
\lbl{mot}
\en

It can be seen from (\ref{mot}) that the field 
$ \ss_0^{-1}(p)\dd\ss(p,q)\ss_0^{-1}(-q) $
depends only on $p+q$ and not by $p$ and $q$ separately.
Therefore, we can identify the local field $\aa$ of Ref.~\cite{PDV} with this
combination of fields:

\eq
\aa(p+q) = \ss_0^{-1}(p)\dd\ss(p,q)\ss_0^{-1}(-q)
\lbl{alpha_0}
\en

Substituting (\ref{alpha_0}) into (\ref{mot}) we can eliminate $\aa$
and we immediately get eq. (\ref{con}) (after the introduction of a Feynman 
parameter and the integration over $k$).\\

The above discussion implies that the bilocal approach that is essential 
to treat large $N$ theories with matter transforming according to the 
fundamental representation of the gauge group ( vector-like matter) 
reduces in the limit of zero gauge coupling constant to the standard large 
$N$ approach used in the pure vector models.

Finally it can be easily seen that a complete potential of the form (\ref{pot})
has just the effect to change the equation for the renormalization
of mass (\ref{ren}):
\eq
\dd m^2 \equiv g^2 \int \ftt{2}{p}\ss_0(p) + \sum_{k=2}^M {\bar c_k\over (k-1)!}
\left(\int \ftt{2}{p}\ss_0(p)\right)^{k-1}
\en
Here we have to define a new renormalized scalar self-coupling to be used in 
the equations for the fluctuations
\eq
v_R \equiv V''\left(\int \ss_0\right) 
= \sum_{k=2}^M {\bar c_k\over (k-2)!}
\left(\int \ftt{2}{p}\ss_0(p)\right)^{k-2}
\en
This procedure is in agreement with the one presented in Ref.~\cite{PDV}.

From the above discussion, we have seen that the bilocal field formalism is
a more general method than the one based on a local composite, that 
can be applied both to the pure vector models and to models containing a 
gauge field whose number of components goes to $\infty$ as $N 
\rightarrow \infty$.

\subsection{A $CP^{N-1}$-like model in bilocal field approach}

The model that we will discuss in this subsection is a massless scalar field,
in the fundamental representation of the $SU(N)$ gauge group,
minimally coupled to a gauge field in adjoint representation and 
subject to the constraint $ \fd_i (x) \ff_i (x) = N/2f $.
The main difference with respect to 
the $CP^{N-1}$ model is the presence of the kinetic term for the gauge 
field and the fact that the gauge field is not abelian, so that its number of 
components grows with $N$.

We therefore start with the model
\eq
\L= -{1\over 4} F^a_{\mu\nu}F^{a\mu\nu} 
    + (D_\mu\ff)^\dagger (D^\mu\ff)
\en
The potential for the scalar field is replaced by the constraint
\eq
\fd_i(x)\ff_i(x)=N/2f_0
\lbl{constraint}
\en
where $f_0$ is a dimensionless constant.
We can express the constraint (\ref{constraint}) by introducing a 
Lagrange multiplier in the action, so that the partition function will
take the form
\eq
\Z = \int \D\ss\D\ll {\rm e}^{iS[\ss]-i\int d^2x \ll_x(\ss^{xx}-N/2f_0 )}
\en
Where now $\ll_x$ is a (local-field) Lagrange multiplier not to be 
confused with the constant previously used in the definition of $1/k_-^2$.
After the usual rescaling $\ss\to N\ss$ we can obtain an action 
$S[\ll,\ss]$ that can be solved in the large-$N$ limit via a saddle point
approximation.\\
The saddle point is obtained by imposing the {\it two} conditions:
\eqa
&& {\dd S\over \dd\ss }\biggr\vert_{\ss=\ss_0, \lambda=\lambda_0} = 0 
\lbl{ss-eq}
\\
&& {\dd S\over \dd\ll }\biggr\vert_{\ll=\ll_0, \sigma = \sigma_0} = 0 
\lbl{l-eq}
\ena
They imply the following equations:
the one for $\ss$
\eq
i\ss_0^{-1}(p,q)= 
p^2(2\pi)^2\dd^{(2)}(p+q) - \ll_0(p+q)
-g^2\int\ftt{2}{k}{(2p_-+k_-)^2\over k_-^2}\ss_0(p+k)  
\en
and 
\eq
\int \ftt{2}{p_1}\ss_0(p_1,l-p_1)={1\over 2f_0}\, (2\pi)^2\dd^{(2)}(l)
\en

The main difference between (\ref{ss-eq}) and (\ref{l-eq}) is that 
$\ll$, being a {\it local} field, has a constant  vacuum expectation 
value. So we can write 
\makebox{$\ll_0(l) = \bar \ll\cdot(2\pi)^2\dd^{(2)}(l)$} and the
previous equations become

\eq
\GG(p) = -g^2\int \ftt{2}{k} {(2p_-+k_-)^2-k_-^2 \over k_-^2}
      {i\over (p+k)^2-m^2+\GG(p+k)+i\ee}
\lbl{g2-eq}
\en
and
\eq
{1\over 2f_0} = \int \ftt{2}{p} {i\over p^2-m^2+\GG(p)+i\ee}
\lbl{gap-eq}
\en
or
\eq
{2\pi\over f}=\log\left(\mu_0^2\over m^2-g^2/\pi\right)
\lbl{f-ren}
\en
with the usual {\it ansatz} (\ref{ss0}) for $\ss_0$, and with
$m^2 \equiv \bar \ll + \dd m^2$. We needed a renormalization of $f$ 
($2\pi/f_0 = 2\pi/f + \log(\LL^2/\mu_0^2)$) and
$\mu_0^2$ is the renormalization scale. Eq.
(\ref{f-ren}) is expression of the asymptotic freedom of the model, because
the coupling constant goes to zero as the scale $\mu_0^2$ goes to infinity.
Comparing (\ref{gap-eq}) with (40) of~\cite{CPN}, one gets some similarities;
in fact, the only difference is the replacement of the mass parameter $m^2$
with $m^2-g^2/\pi$: even in this gauge-interacting (and then nonlocal) version,
the dependence on the cutoff of the coupling $f$ remains the same, and so
it happens for his `asymptotic freedom' property.
\\
We then can see how the model with a potential is related to the one with a 
constraint. In both cases we get the same equation ( compare eq.
(\ref{g2-eq}) with eq. (\ref{gamma-eq})), but in one case the bare squared
mass $m_0^2$ enters directly in the Lagrangian while in the other case 
is generated by the $v.e.v.$ of the field $\ll_x$. When the kinetic term 
for the gauge field is absent ( as in the real $CP^{N-1}$ model ) 
we simply have the {\it gap-equation} 
(\ref{gap-eq}), and the $v.e.v$ of the field $\ll$ 
plays the role of the physical mass $m^2$~\cite{CPN}.
We can now write the equation for the spectrum. It can be 
obtained by writing small fluctuations around the master field 
$\ss = \ss_0 + {1\over \sqrt{N}}\dd\ss$, 
$\ll = \ll_0 + {1\over \sqrt{N}}\dd\ll$.  The equations that one obtains 
are practically identical to (\ref{coupled}), 
apart a term $ \dd\ss_{xx}\dd\ll_x $ in coordinate space, that ensures
$\dd\ss_{xx}=0$. It constrains the integral of the $\vf$ field to be zero.
The eqs. of motion for the fluctuations are (see eq. (\ref{coupled}))

\eqa
 \left[ r^2 - { M^2\over x(1-x)} \right] \vf(x) = 
  &-& {g^2\over \pi} P \int_0^1 {dy\over (y-x)^2}
{(x+y)(2-x-y)\over 2x\, 2(1-x)}\, \vf(y)
\nonumber\\
 &-& {\dd\ll\over 2x(1-x)} 
\ena
and
\eq
\int_0^1 dy \,\vf(y) = 0
\en

If we define
\eq
F(m,r^2) \equiv P \int_0^1 {dx\over r^2\, x(1-x)-m^2 } 
\lbl{fdef}
\en
we get, for $g=0$, the equation $F(m,r^2)=0$, that cannot be  
satisfied unless $r^2=\infty$.
If $g\neq 0 $ we can just use the constraint $\int_{0}^{1} dx \, \vf(x) = 0$ 
to obtain $ \dd\ll $;
using this equation back in the original equation we get
\eqa
 \left[ r^2 - { M^2\over x(1-x)} \right] \vf(x) = 
  &-& {g^2\over \pi} P \int_0^1 {dy\over (y-x)^2}
{(x+y)(2-x-y)\over 2x\, 2(1-x)}\, \vf(y)
\nonumber\\
 &+& {\U_\vf (M,r^2) \over x(1-x)} 
\ena
with 
\eq
 \U_\vf (M,r^2) \equiv {g^2\over 4\pi}{1\over F(M,r^2)}
     P \int_0^1 {dxdy\over r^2\, x(1-x)-M^2 } 
       {(x+y)(2-x-y) \over (y-x)^2 }\, \vf(y)
\en

This equation is very similar to the one in (\ref{coupled}), where now 
the term with $\U_\vf $ takes the place of the last term with 
$ { v\over 4\pi}$ in eq. (\ref{coupled}).

\vfill\eject
\setcounter{equation}{0}
\section{QCD2 with quartic fermion interaction}

We can add to the Lagrangian of $QCD_2$ a term quadratic in
the fermion density $\rr$ without destroying the renormalizability of 
the theory. The request of Lorentz invariance leaves us only three terms: 
$(\pb\pp)^2$ , $(\pb\gg_5\pp)^2$ and $(\pb\gg_\mu\pp)^2$. The resulting 
effective action will have a local term that is not linear in the 
$\rr$-densities. 

The model is  described by the following Lagrangian:

\eqa
\L &=& -{1\over 4} F^a_{\mu\nu}F^{a\mu\nu}
+i\pb\Dir\pp - m^{(1)}_0\pb\pp - m^{(2)}_0\pb\gg_5\pp
\nonumber\\ 
&-& f_1(\pb\pp)^2 - f_2(\pb\gg_5\pp)^2 
- f_3(\pb\gg_\mu\pp)^2  
\lbl{qcd_nl}
\ena
where the $f_i$ are three dimensionless coupling constants.
In the light-cone gauge, we can eliminate the gauge field $A^a_-$ to get a
nonlocal term in the action

\eqa
S &=& \int d^2x d^2y \left\{
\left[i\pb\Dir\pp - m^{(1)}_0\pb\pp - m^{(2)}_0\pb\gg_5\pp
- f_1(\pb\pp)^2\right. \right. 
\nonumber\\
&-& \left. \left. f_2(\pb\gg_5\pp)^2 - f_3(\pb\gg_\mu\pp)^2 \right]\dd_{xy} 
+ {1\over 2} g^2 J^a_- {1\over \partial^2_-}J^a_- \right\}
\ena

We can write the effective action in the bilocal field formalism using as
master field the (bilocal) matrix~\cite{MC}
\eq
U = \mat{\rr_R}{\rr_-}{\rr_+}{\rr_L}
\en
We get (see~\cite{DAS} for the calculation of the Jacobian) 
\eqa
S_{eff}/N &=&  Tr(DU+i\log U)
+ {g^2\over 2}\int d^2x d^2y G_{xy}U^{12}_{xy}U^{12}_{yx}
\nonumber\\
&-& \int d^2x d^2y \left\{
 {a\over 2}\left[ \left(U^{11}_{xy}\right)^2 +\left(U^{22}_{xy}\right)^2 \right]
+  b U^{12}_{xy}U^{21}_{xy} + c U^{11}_{xy}U^{22}_{xy} \right\}\dd_{xy}
\nonumber\\
\ena
where

\eq
D(x,y) = \dd_{xy}
\mat{-{m^{(1)}_0+m^{(2)}_0\over \sq2}}{i\partial_-}{i\partial_+}
{-{m^{(1)}_0-m^{(2)}_0\over \sq2}}
\en
and $a=(f_1+f_2)N$ , $b=2f_3N$, $c=(f_1-f_2)N$ are kept fixed as $N\to\infty$.
The theory is chiral invariant if $m_0 =a=0$.

We can move to Fourier space, and if we impose the saddle-point condition
\eq
{\dd S \over U^{ji}(-q,-p)}\biggr\vert_{U=U_0} = 0 
\en
we get
\eqa
0 &=& \left(D+{i\over U_0}\right)^{ij}(p,q)
+ g^2\dd^{j1}\dd^{i2}\int\ftt{2}{k} {1\over k_-^2} U_0^{12}(k+p,-k+q)
\nonumber\\
&-&\int \ftt{2}{p_1}\ftt{2}{q_1} M_0^{ij}(p_1,q_1)
(2\pi)^2\dd^{(2)}(p_1+q_1-p-q)
\ena
with
\eq
M_0=\mat{a U_0^{11} + c U_0^{22}}{b U_0^{12}}{b U_0^{21}}{a U_0^{22} 
+ c U_0^{11}}
\lbl{m0def}
\en
Notice that, if $a= b+c =0$ $M_0$ is proportional to $ U_{0}^{-1}$.

Multiplying both sides of the saddle point equation with $U_0^{jl}(p,q)$ and 
choosing the usual translationally invariant {\it ansatz} 
\makebox{$ M_0 (p,q) = (2 \pi )^2 \delta ( p+q) M_0 (p) $} we obtain
\eq
\left((D-A)U_0\right)^{il}(p) + i \dd^{il}
+ g^2 \dd^{i2}\int \ftt{2}{k} U_0^{12}(p+k)U_0^{1l}(p)=0
\lbl{sp}
\en
where the matrix $A$ is given by 
\eq
A^{ij} = \int \ftt{2}{k} M_0^{ij}(k)
\lbl{gap_nl}
\en
If one compares eq. (\ref{sp}) with eq.(13) of Ref.~\cite{MC} one can see that 
the only difference  between the two is the shift:

\eq
          \left\{ 
                     \begin{array}{cc}
                     m_{0L}  \rightarrow m_{0L}~   +  ~ A^{11}\\
                     p_-~~  \rightarrow p_-~~~ -  ~ A^{12}\\
                     p_+~~  \rightarrow p_+~~~ -  ~ A^{21}\\
                     m_{0R}  \rightarrow m_{0R}~   +  ~ A^{22}\\
                     \end{array}
                     \right.
\lbl{shift}
\en          
with $m_{0L},m_{0R}=(m^{(1)}_0\mp m^{(2)}_0)/\sq2 $. 
The solution of the saddle point equation is given by
\eq
U_0(p)=
-2D_F(p')\mat{-m_{0L} - A^{11}}{p~'_-}{p~'_+  +{1\over 2p~'_-}\GG(p~'_-)}
{-m_{0R} - A^{22}}
\en
with
\eq
D_F(p') = 
{i \over 2p~'_+p~'_- 
-(m^{(1)}_0-m^{(2)}_0+\sq2 A^{11})
 (m^{(1)}_0+m^{(2)}_0+\sq2 A^{22}) +\GG(p~'_-)+i\ee}
\en
\eq
\GG(p_-) = {g^2\over \pi} - {g^2\over\pi}{|p_-|\over\lambda}
\en
We have changed the definition of $\GG(p)$ with respect to Refs.~\cite{tH2} 
and~\cite{MC}, so that we have replaced
$\GG(p_-)\leftrightarrow -p_-\GG(p_-)$; the definition of the shifted momenta is
$ p~'_\pm \equiv (p-\bar a)_\pm $ with
$\bar a_+ = A^{21}$ , $\bar a_-=A^{12}$ (see eq. (\ref{shift})).

Now we have to enforce eq. (\ref{gap_nl}) as consistency condition.
A simple calculation shows that 
$(A^{22}+A^{11})/2\equiv \dd m^{(1)}/\sq2 $, 
$(A^{22}-A^{11})/2\equiv \dd m^{(2)}/\sq2 $, with
\eq
\dd m^{(1)} = (a+c)\int \ftt{2}{p}{2i~m_1\over
    2p_+p_- -m^2+\GG(p_-)+i\ee}
\en
\eq
\dd m^{(2)} = (a-c)\int \ftt{2}{p}{2i~m_2\over
    2p_+p_- -m^2+\GG(p_-)+i\ee}
\en
and we have defined the renormalized
masses as $m_i\equiv m^{(i)}_0+\dd m^{(i)} $, and 
$m^2\equiv m_1^2 - m_2^2$,
so that we get the dependence of the coupling constants
on the renormalization scale $\mu_0^2$:

\eqa
{2\pi \over a+c}  &=& 
{m_1\over\dd m^{(1)}}\log\left(\mu_0^2\over m^2-g^2/\pi\right)
\nonumber\\
{2\pi \over a-c}   &=& 
{m_2\over\dd m^{(2)}}\log\left(\mu_0^2\over m^2-g^2/\pi\right)
\lbl{gv-ren}
\ena

A similar behaviour was 
already known in the context of the study of the $CP^{N-1}$ 
model with quarks, (see eq. (57) of~\cite{CPN2}, for instance), and as 
in eq. (\ref{f-ren}) the only difference, due to the gauge kinetic term, is to
shift the squared mass of an amount $g^2/\pi$\footnote{
It is important to notice that in the case of $a=c$ and $m^{(i)}_0=0$,
eq. (\ref{gv-ren}) gives the scale of the spontaneous symmetry breaking
$$\dd m^2 = g^2/\pi + \mu_0^2 \exp\left(-{2\pi\over a+c}\right)$$ 
We have a difference with respect to the analysis made in~\cite{GN}
due to the gauge shift $g^2/\pi$ that appears in the correction $\dd m^2$
but that cancels out in the shifted mass 
\makebox{$M^2 = m^2-g^2/\pi = \dd m^2 - g^2/\pi$} 
giving substantially the same result.}.
\\
Furthermore, a careful analysis shows that after regularization 
\makebox{$A^{12}=A^{21}=0$,} due to the fact that $U^{12}(p)$ is odd in $p$.

In conclusion, the only effect on the master field of the interaction terms
is to renormalize the mass, just as in the Bosonic theory and $U_0$ is 
given by 
\eq
U_0(p)=
{-2i \over 2p_+p_- -m^2 +\GG(p_-)+i\ee}
\mat{-{m_1-m_2\over\sq2}}{p_-}{p_+  +{1\over 2p_-}\GG(p_-)}
{-{m_1+m_2\over \sq2}}
\en
The equation for the fluctuation can be obtained as in~\cite{MC}, where,
using this formalism, we solved $QCD_2$ with many flavors and
with a chiral mass term $m^{(1)}\pb\pp+m^{(2)}\pb\gg_5\pp$.
For the sake of simplicity we choose $a=c$ and $m^{(2)}=0$,
so that $m_L=m_R=m/\sq2$.
Defining
\eq
\C^{ij}(p,-q) = \mat{- {m/\sq2\over |q_-|}}{1}{{m^2/2 \over 
|p_-||q_-|}}{- {m/\sq2\over |p_-|}}
\en
the form of the resulting equations suggests the following {\it ansatz} for the 
solutions
\eq
\vf^{ij}(r,s_-) = 
\th\left(\rmd-|s_-|\right)\C^{ij}\left(s_-+\rmd,s_--\rmd\right)\vf(s_-) 
\en
so that finally we get after the usual rescaling of variables 
the correspondent of the 't Hooft equation for $QCD_2$ 
with quartic fermion interaction:

\eqa
&&\left[ r^2 - {M^2\over x(1-x)} \right]\vf(x) = -{g^2\over\pi}
P\int_0^1 {dy\over (y-x)^2} \vf(y) 
\nonumber\\
&&+ {a+c\over 2\pi} m^2\left( {1\over x}\int_0^1 {\vf(y) \over 1-y}dy 
             + {1\over 1-x}\int_0^1 {\vf(y) \over y}dy\right)
\nonumber\\
&&+ \bdpi m^2 \left( {1\over x(1-x)}\int_0^1 \vf(y)dy 
             + \int_0^1 {\vf(y) \over y(1-y)}dy\right)
\lbl{bound_quartic}
\ena 
As in section 3, we can write the solution in the case $b\ne 0$, $a+c\ne 0$, 
$g=0$.
The previous equation becomes (we rescale $r^2 = m^2\mu^2$) 

\eqa
&&\left[ \mu^2 - {1\over x(1-x)} \right]\vf(x) = 
\nonumber\\
&&+ {a+c\over 2\pi} \left( {1\over x}\int_0^1 {\vf(y) \over 1-y}dy 
             + {1\over 1-x}\int_0^1 {\vf(y) \over y}dy\right)
\nonumber\\
&&+ \bdpi \left( {1\over x(1-x)}\int_0^1 \vf(y)dy 
             + \int_0^1 {\vf(y) \over y(1-y)}dy\right)
\lbl{phi_thirring}
\ena 
that gives
\eq
\vf(x) = \bdpi {A + \rr B + 2B\,x(1-x)\over \mu^2\,x(1-x) - 1}
\label{aaaa}
\en
where $A \equiv \int \phi(x) $, $B \equiv \int \phi(x)/x $, $\rr = (a+c)/b$.
This solution is remarkably similar to the one obtained by Fujita and Ogura
~\cite{Fujita} for the massive Thirring model~\cite{Thirring} with $N_f=1$ ; 
indeed, due to the equality
\eq
(\gg_\mu)_{ab}(\gg^\mu)_{cd} = \uno_{ad}\uno_{bc} 
- (\gg_5)_{ad}(\gg_5)_{bc} 
\en
in the case of $N_f=1$ our starting Lagrangian and the one
of the Thirring model coincide for $a=c = b/2 = g^*/4$ where $g^*$ is the
Thirring coupling constant 
(there is a different sign with respect to ~\cite{Fujita} 
due to the anticommutation of the fermions caused
by $\Sigma_a T^a_{ij}T^a_{kl}$ 
and a factor of 2 that follows from our definition of the fermionic densities
$\rr \sim \sq2 \pb\pp$).

As in~\cite{Fujita} from eq.~(\ref{aaaa})  we obtain the two  equations
\eqa
A &=& \bdpi\left[F(\mu^2)\left\{A + \left(1+{2\over \mu^2}\right)B\right\}
    +{2\over \mu^2}B\right]
\nonumber\\
B &=& \bdpi\left[{\mu^2\over 2}F(\mu^2)\left\{A 
+ \left(1+{2\over \mu^2}\right)B\right\}
    +(A+B)\log\ee\right]
\lbl{matrix}
\ena
where $\ee$ is a cutoff needed to regularize the divergent equations, and
\par\noindent
\makebox{$F(\mu^2) \equiv F(1,r^2/m^2)$} (see (\ref{con})). 

We must renormalize (\ref{matrix}) by adding a mass counterterm that verifies
\eq
\dd\mu^2\, \vf(x) = \bdpi\left[\ee_1 + {\ee_2\over x(1-x)}\right]
\lbl{dmass}
\en
Adding (\ref{phi_thirring}) to (\ref{dmass}) we get
\eq
\vf(x) = \bdpi {A + \rr B + \ee_2 + (2B+\ee_1)\,x(1-x)\over \mu_R^2\,x(1-x) - 1}
\en
where $\mu_R^2 \equiv \mu^2 + \dd\mu^2$. In the following we will omit
the suffix $R$ for short.

From (\ref{matrix}) we observe that choosing 
\makebox{$\ee_1 = -b/2\pi\cdot 2(A+B^*)\log\ee$},\par\noindent
\makebox{$\ee_2 = -b/2\pi\cdot\rr\,(A+B^*)\log\ee$}, with 
\makebox{$ B^* = B - b/2\pi\cdot(A+B^*)\log\ee$} we get
\eq
\vf(x) = \bdpi {A + \rr B^* + 2B^*\,x(1-x)\over \mu^2\,x(1-x) - 1}
\en
Now both $A$ and $B^*$ are {\it finite}, so we can solve the equations
\eqa
A &=& \bdpi\left[F(\mu^2)\left\{A + \left(\rr+{2\over \mu^2}\right)B^*
\right\}+{2\over \mu^2}B^*\right]
\\
B^* &=& \bdpi\,
 {\mu^2\over 2} F(\mu^2)\left[A + \left(\rr+{2\over \mu^2}\right)B^*\right]
\nonumber
\lbl{matrix2}
\ena
getting 
\eq
{b\over \pi}\left[\rr\,{\mu^2\over 4} + 1 + {b\over 4\pi}\right]
F(\mu^2) = 1
\lbl{bound_th}
\en
Let us first examine the case $\rr=1$.
We can find bound states if (\ref{bound_th}) has a solution for $0<\mu^2<4$.
With the usual parametrization $\mu = 2\,sin\th$, $0<\th<\pi/2$, 
we have to solve the
equation ($\ll=-b/2\pi$)
\eq
\ll^2-\ll\,(3-\cos\,2\th) + {\sin\,2\th\over 2\th}=0
\en
that, solved with respect to $\ll$, gives
\eq
2\ll = 3-\cos\,2\th \pm \sqrt{(3-\cos\,2\th)^2-4\,{\sin\,2\th\over 2\th}}
\en
When $\th$ varies from 0 to $\pi/2$, $\ll_-$ goes from 1 to 0, whereas
$\ll_+$ goes from 4 to 1; this gives
the allowed range for the coupling constant in order to have bound states:
$ 0<\ll<4$. In Fig.3 we plot the two branches of $2\ll(\th)$, and one can note
the singularity at $\ll=1$ of the function $\th(\ll)$, that will be analyzed
later.
If $\ll=1~~(b=-2\pi)$ we have $\mu^2=0$ i.e. at the singularity we find the well
known massless pseudoscalar bound state of the massive Thirring 
model~\cite{Aoyama}, that corresponds to the fundamental boson of the 
sine-Gordon theory~\cite{Dashen}.

Finally, we want to investigate the presence of tachyons in our model 
by searching a solution of (\ref{bound_th}) in the range
$\mu^2<0$. Putting $\mu^2 = -4\nu^2$ we get
\eq
\ll^2 - 2\ll\,(1-\nu^2) + {\nu\sqrt{1+\nu^2}\over\log(\nu+\sqrt{1+\nu^2})}=0
\lbl{tac}
\en
The equation has solutions iff $\ll<0$ (see Fig.4); at $\ll=1-\nu_0^2$ where 
\makebox{
$\nu_0\approx 1.59803 $} we have the 'lightest'
tachyon, with mass $\mu^2_0 = -4\nu_0^2$; for any value of
$\ll>0$ eq. (\ref{tac}) doesn't have any solution, so we conclude
that $\ll>0$ is the physical range of the coupling constant.
In the tachyonic sector we have the following limits:
\eqa
\mu^2(\ll) &\approx& -{\rm e}^{-{1\over \ll}}~~~\ll\to 0^-
\nonumber\\
\mu^2(\ll) &\approx& 2 \ll~~~~~\ll\to -\infty
\ena

For a generic $\rr$ we have 
\eq
\ll^2-\ll\,[2+\rr\,(1-\cos\,2\th)] + {\sin\,2\th\over 2\th}=0
\en
that, solved with respect to $\ll$, gives
\eq
2\ll = 2+\rr-\rr\,\cos\,2\th \pm 
\sqrt{(2+\rr-\rr\,\cos\,2\th)^2-4\,{\sin\,2\th\over 2\th}}
\lbl{llrho}
\en
Now the allowed range of $\ll$ goes from $1+\rr-|1+\rr|$ to
$1+\rr+|1+\rr|$: when $0<\th<\pi/2$,  $\ll$ is always positive if $\rr>-1$, 
and always negative if $\rr<-1$.
We can also see that for any $\rr$ $\th=0$ implies $\ll=1$ i.e.
at the critical point $\ll = 1$ ($b=-2\pi$) we have a massless bound state.

Furthermore, if $\rr<\rr_c=-1/3$ 
the function $\th(\ll)$ has a turning point at $\th=\th_0>0$, 
as it can be seen from the expression of the
argument of the square root in (\ref{llrho}) for small $\th$:
\eq
(2+\rr-\rr\,\cos\,2\th)^2-4\,{\sin\,2\th\over 2\th} 
\approx (2/3+2\rr)\,4\th^2 
\en
The behaviour of the argument of the square root in (\ref{llrho}) is shown in 
Fig.5. We have forbidden values of $\th$ if the function is negative.
The value of the turning point $\th_0$ increases for negative $\rr$ and
covers the whole range of $\th$ at $\rr=-1$ (Fig.6); after that, the gap
decreases in extension. The point $\rr=-1$ is very
important, because it is the supersymmetric point in the full theory
with fermions coupled to bosons~\cite{CPN2}. At $\rr=-1$ we can only have the
massless bound state at the critical point $\ll=1$, because 
(\ref{llrho}) has no real solutions except $\ll=1$ ($\th=0$).

We can also see that near the value $\ll=1$ the angle $\th$ has the behaviour
\eq
\th(\ll)\approx \syst
 {{3\sqrt{5}\over \sqrt{3+\sqrt{5}}}\sqrt{1-\ll}~~~~\ll\to 1^-}
 {{3\sqrt{5}\over \sqrt{3-\sqrt{5}}}\sqrt{\ll-1}~~~~\ll\to 1^+}
\en 
if $\rr= -1/3$, and
\eq
\th(\ll)\approx \sqrt{{1\over 2\rr+2/3}}\, |1-\ll|~~~\ll\to 1
\en 
if $\rr> -1/3$, that shows the transition at $\rr=-1/3$ and the singular 
behaviour of $\th(\ll)$ near $\ll=1$.

We can see more clearly what's happening 
if we investigate the tachyonic sector.
The coupling constant is given by
\eq
\ll = 1-\rr\nu^2\pm
\sqrt{(1-\rr\nu^2)^2-{\nu\sqrt{1+\nu^2}\over \log(\nu+\sqrt{1+\nu^2})}}
\lbl{llrhotac}
\en

In Fig.7 we can see the behaviour of the argument of the square root in 
(\ref{llrhotac}): the function $\nu(\ll)$ has a turning point $\nu_0>0$
(for example \makebox{$\nu_0\approx 1.59803$} if $\rr=1$) that
disappears at $\rr\le\rr_c=-1/3$, the critical point seen before.

The special case $\rr=0$ contains no tachyons for any value of $\ll$, since
it can be seen that (\ref{llrhotac}) has no real solutions with nonzero $\nu$.

Thus, it has turned out that in the tachyonic region
$\ll$ has always the opposite sign of $\rr$: 
we have tachyons in our theory if $\rr\ll<0$ ; since
$\ll=-b/2\pi$, $\rr=(a+c)/b$ we can conclude that the presence of tachyons
depends only on $a+c$ : {\it we have tachions in the theory if $a+c>0$}.
Our theory of self-coupled fermions is
well defined provided that the coefficient of $(\pb\pp)^2$ in our
starting lagrangian (\ref{qcd_nl}) is positive (i.e. $f_1<0$).\\

In conclusion, in the free-tachyon sector we have a bound state
for \par\noindent
\makebox{$2(1+\rr)<\ll<0$} if $\rr<-1$ and for 
\makebox{$0<\ll<2(1+\rr)$} if $\rr\ge 0$.
The mass of the bound state is given by $M=2m_F\sin\th(\ll)$,
with $\th(\ll)$ given by (\ref{llrho}).
If $\rr=-1$ the only allowed value for the coupling constant is $\ll=1$,
for which the tachyon disappears and we only have a massless bound state;  
in the range $-1<\rr<0$ we cannot have bound states without having also 
tachyons, so we won't examine this range of the parameter $\rr$ 
(and the point $\rr_c=-1/3$) far in detail.

We finally observe that an 
interesting case is $b=0$, $a+c=2f_1N\ne 0$, that corresponds
to the Gross-Neveu model~\cite{GN} if one identifies $f_1 \leftrightarrow 
-g^2/2$, $g^2$ being the Gross-Neveu coupling constant (not to be confused 
with the gauge coupling), as used in~\cite{GN}. One can see that the previous 
calculations lead to 
\eq
1={a+c\over 4\pi}\,\mu^2 F(\mu)
\lbl{Gross-Neveu}
\en
that admits a bound state (note that such a bound state is never massless)
if $a+c<0$, given by the solution of ($\mu=2\sin\th$)
\eq
1 = -{a+c\over \pi}\,\th\,\tan\th 
\en
If $a+c>0$ equation (\ref{Gross-Neveu}) has a solution in the tachyonic sector 
and the theory does not admit a ground state~\cite{GN}~\cite{Feldman}.
This confirms the results obtained before.

\vfill\eject
\setcounter{equation}{0}
\section{Complete gauge theory}

Finally, now that we've got experienced in using this formalism, we
are ready to solve the full gauge theory 
(\ref{s-prime}):
\eqa
\L= &-&{1\over 4} F^a_{\mu\nu}F^{a\mu\nu}
+\pb(i\Dir)\pp + (D_\mu\ff)^\dagger (D^\mu\ff)-\tilde V(\fd\ff)
-\tilde W_1(\fd\ff)\pb\pp 
\nonumber\\
&-&\tilde W_2(\fd\ff)\pb\gg_5\pp
-F_1(\fd\ff)(\pb\pp)^2 - F_2(\fd\ff)(\pb\gg_5\pp)^2 
- F_3(\fd\ff)(\pb\gg_\mu\pp)^2
\nonumber\\
\lbl{s-p2}
\ena
where we require the potentials to be polynomials in $\fd\ff$.
Using the light cone gauge we are able to remove $A^a_\mu$ from the Lagrangian:
\eqa
\L &=&\pb(i\Dsl)\pp + \fd(-\partial^2)\ff-\tilde V(\fd\ff)
-\tilde W_1(\fd\ff)\pb\pp -\tilde W_2(\fd\ff)\pb\gg_5\pp
\nonumber\\
&-& F_1(\fd\ff)(\pb\pp)^2 - F_2(\fd\ff)(\pb\gg_5\pp)^2 
- F_3(\fd\ff)(\pb\gg_\mu\pp)^2+ {1\over 2} g^2 J^a_- {1\over \partial_-^2} J^a_-
\nonumber\\
\lbl{s-p3}
\ena

This is our (nonlocal) starting Lagrangian, with $J^a_-$ 
that is given by (\ref{ja}):
\eq
J^a_- = \pb T^a \gamma_- \pp + i \fd T^a \ddm \ff
\en

\subsection{The 't Hooft equation for fermion-boson bound states}

We begin discussing the case 
\eqa
&&\tilde V(\fd\ff) = m^2_{0B}\fd\ff
\nonumber\\
&&\tilde W_1(\fd\ff) = m_{0F}
\\
&&\tilde W_2(\fd\ff) = 0
\nonumber\\
&& F_i(\fd\ff)=0
\nonumber
\ena
\\
First, we eliminate the unnecessary degree of freedom $\psi_L$ from
the Lagrangian (its kinetic term only contain $\part_-$), and in the 
following we will define \makebox{$\pp\equiv2^{1\over 4} \pp_R$} 
(this is to remove boring $\sq2$ terms out the 
Lagrangian); then, to rewrite the whole interaction Lagrangian, we have 
to define four bilocal fields:

\eq
\rho^{xy} \equiv \sum_A \pb_A(y)\pp_A(x)
\en
\eq
\ss^{xy} \equiv \sum_A \fd_A(y)\ff_A(x)
\en
\eq
\cc^{xy} \equiv \sum_A \fd_A(y)\pp_A(x)
\en
\eq
\cb^{xy} \equiv \sum_A \pb_A(y)\ff_A(x)
\en
In terms of the previous bilocal the interaction Lagrangian becomes:
\eq
\L_{int}={ g^2\over 2}G_{xy}
(\rho^{yx}\rho^{xy} + \ss^{yx}i\ddu i\ddd\ss^{xy}
 -\cc^{yx}i\ddu\cb^{xy} + \cb^{yx}i\ddd\cc^{xy})
\en
Now, it is a bit more difficult to get the Jacobian of the change of variables
$$ \D\pb\D\pp\D\fd\D\ff \rightarrow \D\rho\D\ss\D\cb\D\cc $$ 
it is expressed by:
\eq
J[\rho,\ss,\cb,\cc]=\int \D\pb\D\pp\D\fd\D\ff\, \dd(\rho-\pb\pp)\dd(\ss-\fd\ff)
                             \dd(\cb-\pb\ff)\dd(\cc-\fd\pp)
\en
By using auxiliary fields to exponentiate the $\dd$'s, we can calculate 
the Jacobian in the large-N limit:
\eq
J=exp(-N\, STr\log U)
\en
with
\eq
U=\mat{\rho^{xy}}{-\cc^{xy}}{\cb^{xy}}{-\ss^{xy}}
\lbl{u-def}
\en
$STr$ is the Supertrace, defined as
\eq
STr\mat{m_1}{\mu_2}{\mu_1}{m_2} \equiv Tr (m1) - Tr (m2)
\en
with $m_i$ and $\mu_i$ that are matrices with commuting and anticommuting
elements, respectively. This explains the choice of the $-$ signs in 
(\ref{u-def}).\\
By rescaling all fields by $N$, keeping $g^2 N=g^2$ constant we get
the full action of the model:
\eqa
S/N & = &  Tr_{pq}STr(DU+i\log U) 
\nonumber\\
& + & {1\over 2}g^2 \int \ftt{2}{p_1}
      \ftt{2}{p_2}\ftt{2}{k} {c_{ij}(p_1,p_2,k) \over k_-^2}
               U^{ij}(p_1,p_2) U^{ji}(k-p_2,-k-p_1)
\nonumber\\
\lbl{fullaction}
\ena
with
\eq
 D^{ij}(p,q) = \mat{p_+ - {m_{0F}^2\over 2p_-}}{0}{0}{p^2-m_{0B}^2}
                (2\pi)^2\dd^{(2)}(p+q)
\en
\eq
c_{ij}(p,q;k) = \mat{1}{-(2q_--k_-)}{(2p_-+k_-)}{(2p_-+k_-)(2q_--k_-)}
\en
The saddle point equation is obtained by imposing
\eq
{\dd S\over \dd U^{ji}(-q,-p)}\biggr\vert_{U=U_0} = 0 
\lbl{spc}
\en
that gives 

\eq
U_0^{ij}(p,q) = U_0^{ii}(p)\dd^{ij}\dd^{(2)}(p+q) 
\en
with
\eq
U_0^{11}(p) = \rho_0(p)=
 { -2ip_-\over 2p_+p_--M_F^2-{g^2\over\pi}{|p_-|\over\ll}+i\ee}
\en
and (note the sign due to the Supertrace)
\eq
-U_0^{22}(p) = \ss_0(p)=
 { i \over 2p_+p_--M_B^2-{g^2\over\pi}{|p_-|\over\ll}+i\ee}
\en
The equation for the bound states can now be trivially obtained. We 
start from the equation for the fluctuations
\eq
(1-2\dd^{j2})\dd U^{ij}(p,q) = -ig^2 U_0^{ii}(p) U_0^{jj}(-q)
    \int\ftt{2}{k}{c_{ij}(-q,-p;k)\over k_-^2}\dd U^{ij}(k+p,-k+q)
\nonumber\\
\en
that is the generalized equation for the 't Hooft "blob",
which has a leg made of a particle
of type $i$ and the other one made of a particle of type $j$.
The $\dd U^{11}$ component leads to the 't Hooft equation~\cite{tH2}, 
already obtained within this formalism in~\cite{MC}; the $\dd U^{22}$
component gives (\ref{thb}). Finally the other two components 
$\dd U^{12}$ or  $\dd U^{21}$ give the equation for the scalar-fermion bound 
states:
\eq
 r^2 \vf(x) = \left[{M_B^2\over x} + {M_F^2\over (1-x)} \right] \vf(x)
 - {g^2\over \pi} P \int_0^1 {dy\over (y-x)^2}
{(x+y) \over 2x}\, \vf(y)
\lbl{thbf}
\en
recently written by K. Aoki~\cite{Aoki}.

\subsection{Solving the most general Lagrangian}

Our method works also in the case of an arbitrary potential of the form
\eqa
&& 
\tilde V(\ss)=m^2_{0B}\ss+\sum_{n=2} {v_{n}\over n!}{\ss^n\over N^{n-1}}
\nonumber\\
&&
\tilde W_i(\ss)=m^{(i)}_0 + \sum_{n=1}
 {w^{(i)}_{n}\over n!}{\ss^n\over N^{n}} 
\\
&& F_i(\ss)= \sum_{n=0}
 {f^{(i)}_{n}\over n!}{\ss^n\over N^{n+1}}
\lbl{potentials}
\ena

Here the crucial difference is that we cannot remove anymore $\pp_L$ as 
before  and therefore we have to introduce a $3\times3$ bilocal field matrix:
\eq
U=\matre{\rr_R}{\rr_-}{-\cc_R}{\rr_+}{\rr_L}{-\cc_L}{\cb_L}{\cb_R}{-\ss}
\en
As a consequence we get
\eq
D(p,q)=\matre{-{m^{(1)}_0+m^{(2)}_0 \over \sq2}}{~~~~~~p_-}{0}
       {p_+}{-{m^{(1)}_0-m^{(2)}_0 \over \sq2}}{0}
       {0}{~~~~~~0}{p^2-m^2_{0B}}(2\pi)^2\dd^{(2)}(p+q)
\en

and the gauge part now couples the fermionic and the bosonic sectors.

We can also define
\eqa
&& \tilde V(N\ss) = N~\left[m^2_{0B}\ss + V(\ss)\right]
\nonumber\\
&& \tilde W_i(N\ss) = m^{(i)}_0 + W_i(\ss)
\nonumber\\
&& \A(\ss) = N[F_1(N\ss) + F_2(N\ss)]
\\
&& \B(\ss) = 2NF_3(N\ss) 
\nonumber\\
&& \C(\ss) = N[F_1(N\ss) - F_2(N\ss)]
\ena
and the $L/R ~~\rr$-densities are defined (apart a $\sq2$ factor) 
as in~\cite{MC}.

If we impose the saddle point condition (\ref{spc}) to the resulting effective
action with the usual translationally invariant {\it ansatz} we get
(defining as in (\ref{gap_nl}) $A^{ij} \equiv \int M_0^{ij}$)

\eq
(1-2\dd^{j3})\left(D-A+{i\over U_0}\right)^{ij}(p)+g^2\int\ftt{2}{k}
 {H^{ij}(U_0;p,-p;k) \over k_-^2} =0
\lbl{aij2}
\en

The self-interacting part can be written as
\eqa
\V(\ss,\rr)&=&V(\ss) + W_1(\ss)\left({\rr^{11}+\rr^{22}\over \sq2}\right)
+ W_2(\ss)\left({\rr^{11}-\rr^{22}\over \sq2}\right)
\nonumber\\
&+&{1\over 2} \A(\ss)\left[\left(\rr^{11}\right)^2 + \left(\rr^{22}\right)^2\right]
+ \B(\ss)\rr^{12}\rr^{21} + \C(\ss)\rr^{11}\rr^{22}
\lbl{vtot}
\ena

the gauge coupling part is given by the matrix 
$H^{ij}$ that has four nonzero elements:

\eqa
H^{23}(U;p,q;k) &=& (2p+k)U^{13}(k-q,-k-p)
\nonumber\\
H^{21}(U;p,q;k) &=& U^{12}(k-q,-k-p)
\nonumber\\
H^{31}(U;p,q;k) &=& -(2q-k)U^{32}(k-q,-k-p)
\nonumber\\
H^{33}(U;p,q;k) &=& (2p+k)(2q-k)U^{33}(k-q,-k-p)
\nonumber\\
\ena
and the matrix $M_0$ is given by
\eq
M_0 = \matre{{W_1+W_2\over \sq2}+\A\rr_0^{11}+\C\rr_0^{22}}{\B\rr_0^{12}}{0}
            {\B\rr_0^{21}}{{W_1-W_2\over \sq2}+\A\rr_0^{22}+\C\rr_0^{11}}{0}
            {0}{0}{-M_0^{33}}
\lbl{m0def2}
\en
with $M_0^{33}= \dd\V(\ss,\rr)/\dd\ss$.\\
If $U_0$ is translationally invariant it is easy to see that $\int M_0$ is a 
constant. It can be demonstrated that only the diagonal terms of the matrix 
$A^{ij}$ are nonzero, so that we can absorb them in a mass renormalization.
The master field $U_0^{ij}$ is block diagonal:
\eq
U_0 = \left( \begin{array}{clcr}
               {\rr_0^{11}}& {\rr_0^{12}} & {0} \\
               {\rr_0^{21}}& {\rr_0^{22}} & {0} \\
               {0}         & {0}          & {-\ss_0} 
               \end{array} \\
\right)
\en

For $i,j=1,2$ we have

\eq
\rr_0^{ij}(p)= -2D_F(p)
     \mat{-{m_0^{(1)}-m_0^{(2)}\over \sq2}-A^{11}}{p_-}
  {p_++{1\over 2p_-}\GG(p_-)}{-{m_0^{(1)}+m_0^{(2)}\over \sq2}-A^{22}}
\en
with

\eq 
D_F(p) = {i\over 2p_+p_-
     -(m_0^{(1)}-m_0^{(2)}+\sq2 A^{11})(m_0^{(1)}+m_0^{(2)}+\sq2 A^{22})+\GG(p_-)+i\ee}
\en

and

\eq
-U_0^{33}(p) \equiv \ss_0(p) = {i\over 2p_+p_--(m_{0B}^2+A^{33})+\GG(p_-)+i\ee}
\en

The constants $A^{ij}=\int M_0^{ij}$ are defined by using (\ref{m0def2});
one can see that $A^{12}=A^{21}=0$ as before; the remaining constants
must be regularized and then re-absorbed in a mass renormalization:
\eqa
&&\sq2 A^{11} = \dd m^{(1)}-\dd m^{(2)}
\nonumber\\
&&\sq2 A^{22} = \dd m^{(1)}+\dd m^{(2)}
\\
&&A^{33} = \dd m^2_B
\nonumber
\ena
so that, after the introduction of the renormalized masses 
\makebox{$m_B^2 = m_{0B}^2 + \dd m_B^2$ ,} 
\makebox{$ m_i = m_0^{(i)}+\dd m^{(i)} $ ,} 
\makebox{$ m_F^2 = m_1^2-m_2^2 $}
and of the shifted masses \makebox{$M_a^2 = m_a^2-g^2/\pi$,}
\makebox{$a=B,F $}
we can write the propagators in a more familiar form:

\eq
\rr_0^{ij}(p)= 
{-2i\over 2p_+p_-
     -M_F^2-g^2|p_-|/\pi\ll+i\ee}
     \mat{-{m_1-m_2\over \sq2}}{p_-}
         {p_++{1\over 2p_-}\GG(p_-)}{-{m_1+m_2\over \sq2}}
\en

\eq
\ss_0(p) = {i\over 2p_+p_--M_B^2-g^2|p_-|/\pi\ll+i\ee}
\en
\\
It is now easy to get the equations for the fluctuations of the master field,
by applying the same methods as in the previous section. Because of the 
presence of higher than quadratic terms in the densities appearing in 
the lagrangian, as in the $O(N)$ theory, we have to renormalize all the 
constants that couple two densities i.e. the scalar coupling $v_2$, the Yukawa 
couplings $ w^{(1)}_1$ and $w^{(2)}_1$ 
of (\ref{potentials}) and the constants $a_0$ , $b_0$ , $c_0$ that did not need
to be renormalized in the version of the model ($QCD_2$ with quartic 
interaction term) studied in the previous section.
This is our renormalization choice:

\eqa
&& v \equiv   V''(\LL_B) + {1\over 2} \A''(\LL_B)
\left[\left(\LL_F^{11}\right)^2+\left(\LL_F^{22}\right)^2\right]
+ \C''(\LL_B)\LL_F^{11}\LL_F^{22}
\nonumber\\
&& {w_1+w_2\over \sq2} \equiv 
{W_1'(\LL_B) + W_2'(\LL_B)\over \sq2}
+ \A'(\LL_B)\LL_F^{11} + \C'(\LL_B)\LL_F^{22} 
\\
&& {w_1-w_2\over \sq2} \equiv 
{W_1'(\LL_B) - W_2'(\LL_B)\over \sq2}
+ \A'(\LL_B)\LL_F^{22} + \C'(\LL_B)\LL_F^{11} 
\nonumber\\
&& a \equiv \A(\LL_B)~~~~ b \equiv \B(\LL_B) ~~~~
c \equiv \C(\LL_B) 
\nonumber
\ena
Here the prime means derivative with respect to $\ss$, and
we defined three (infinite) constants 
$ \LL_B = \int \ss_0 $ , $ \LL_F^{ii} = \int \rr_0^{ii} $.
\\

With the help of these definitions, we can write the form of the
matrix $\dd\tilde M$, obtained by
expanding the potential (\ref{vtot}) up to second order in the 
fluctuations (compare with (\ref{m0def2})):

\eq
\dd\tilde M =
\matre{a\dd U^{11} + c \dd U^{22} + {w_1 
+ w_2\over\sq2}\dd U^{33}}{b\dd U^{12}}{0}
{b\dd U^{21}}
{a\dd U^{22} + c \dd U^{11} + {w_1 - w_2\over\sq2}\dd U^{33}}{0}
{0}{0}{\dd\tilde M^{33}}
\en
with
\eq
\dd \tilde M^{33} = 
{v\dd U^{33} + {w_1 + w_2\over\sq2}\dd U^{11}
+{w_1 - w_2\over\sq2}\dd U^{22}}
\en

Our final result is the equations for the fluctuations in the most
general case of a 2-dimensional gauge theory:

\eqa 
(1-2\dd^{l3})i \dd U^{kl}(p,q) &=& g^2 U_0^{ki}\left(p\right)
U_0^{jl}\left(-q\right)
\int\ftt{2}{k} {H^{ij}(\dd U;p,q;k)\over k_-^2} 
\nonumber\\
&-&U_0^{ki}\left(p\right)U_0^{jl}\left(-q\right)
 \int\ftt{2}{k}\dd \tilde M^{ij}(k+p,-k+q) 
\nonumber\\
\lbl{fluct6}
\ena

These equations are the generalization of the ones already seen; the effect of 
the interacting potentials is to add inhomogeneous terms in the integral
equations of motion. 
If one writes the entire set of equations, 
one can see that the independent degrees of freedom are the Fermion-Fermion
and Boson-Boson fields $\rr^{12}=\rr_- \equiv\pb_R\pp_R $, 
$\ss\equiv \fd\ff$ and the Fermion-Boson bound state 
$\dd\cc_R\equiv\fd\pp_R$.
Since this last one is not subject to the interactions given by the
additional potentials $V$ , $W_{i}$ 
and $F_i$, it does not develop an inhomogeneous term in his equations of motion.
These are the full equations for the Fermion-Fermion and Boson-Boson bound
states in the case of no chiral $\pb\gg_5\pp$ term 
$m_R=m_L\equiv m_F$, $w_1 = w $, $w_2 = 0 $
(for the equation of the Fermion-Boson bound state, see (\ref{thbf})): 

\eqa
&&\left[ r^2 - {M_F^2\over x(1-x)} \right]\vf_{F}(x) = 
-{g^2\over \pi} P\int_0^1 {dy\over (y-x)^2} \,\vf_{F}(y) 
\nonumber\\
&&+ {a+c\over 2\pi}m_F^2 
\left( {1\over x}\int_0^1 {\vf_{F}(y) \over 1-y}dy 
             + {1\over 1-x}\int_0^1 {\vf_{F}(y) \over y}dy\right)
\nonumber\\
&&+ \bdpi m_F^2\left( {1\over x(1-x)}\int_0^1 \vf_{F}(y)dy 
             + \int_0^1 {\vf_{F}(y) \over y(1-y)}dy\right)
\nonumber\\
&& +{w\over 2 \pi} {m_F\over x(1-x)}\int_0^1 \vf_{B}(y)dy 
\nonumber\\
\ena 
and
\eqa
&&\left[ r^2 - {M_B^2\over x(1-x)} \right]\vf_{B}(x) = 
-{g^2\over \pi} P\int_0^1 {dy\over (y-x)^2}{(x+y)(2-x-y)\over 4x(1-x)}
\,\vf_{B}(y) 
\nonumber\\
&&+ {1\over 4x(1-x)}\left( {v\over \pi}\int_0^1 \vf_{B}(y) dy 
+ {w \over 2\pi} m_F \int _0^1 {\vf_{F}(y)\over y(1-y)}dy
\right)
\nonumber\\
\ena

For $g=0$ we can give the solution of those equations: let us define
\eq
\rr={a+c\over b}~~~~\ss={w\over m_Fb}~~~~\eta={wm_F\over v}
\en
\eq
A=\int_0^1\vf_F(y)dy~~~B=\int_0^1{\vf_F(y)\over y}dy~~~C=\int_0^1\vf_B(y)dy
\en
so that
\eqa
\vf_B(x) &=& {v \over 4\pi}{C+\eta B\over r_B^2\,x(1-x)-m_B^2}
\nonumber\\
\vf_F(x) &=& {b m_F^2\over 2\pi}{A+\rr B+\ss C + 2B\,x(1-x)\over r_F^2\,x(1-x)-m_F^2}
\ena

We then get a new set of equations that after the renormalization of $B$
(see (\ref{matrix})) are given by:
\eqa
A&=&\bdpi\left[F(\mu_F)\left\{A+\ss C +\left(\rr + {2\over \mu_F^2}\right)B^*
\right\}+{2\over \mu_F^2}B^*\right]
\nonumber\\
B^*&=&\bdpi {\mu_F^2\over 2}F(\mu_F)\left\{A+\ss C +\left(\rr + 
{2\over \mu_F^2}\right)B^*\right\}
\nonumber\\
C&=&{v\over 4\pi m_B^2}F(\mu_B)(C+\eta B^*)
\ena
If $1-v/4\pi m_B^2 F(\mu_B)=0$ then $B^*=0$; the only solution that
doesn't imply also $A=C=0$  requires $\mu_F^2=0$ and
\eq
A(1-\ll) = - {w m_F\over 2\pi} C
\en
where we defined $\ll=-b/2\pi$.

If $1-v/4\pi m_B^2 F(\mu_B) \ne 0 $
we can solve the matrix equations by obtaining
\eq
A={2\mu_F^2}(1-\ll)B^*
\en
\eq
\left({\sin\,2\th_B\over 2\th_B}+{v\over 4\pi m_B}\right)C=
-{w m_F\over 4\pi m_B^2}
B^*
\en
\eq
\ll^2-\ll\,[2+\rr\,(1-\cos\,2\th_F)] + {\sin\,2\th_F\over 2\th_F}
+{w\over 2\pi m_F}(1-\cos\,2\th_F)\left({C\over B^*}\right)=0
\en
where we made the usual parametrization $\mu_a=2\sin\,\th_a$,
$0<\th_a<\pi/2$, $a=$F,B.

The equations depend on two constants, that we can fix by normalizing $\vf_F$
and $\vf_B$; requiring $A=C=1$ we get
\eq
1=\left[{v\over 4\pi m_B^2}+{wm_F\over 8\pi m_B^2}{\mu_F^2\over 1-\ll}\right]
F(\mu_B)
\en
\eq
1 = -2\ll F(\mu_F)\left[\rr \,{\mu_F^2\over 4}-
{w\over 4\pi m_F}\left(1-{1\over \ll}\right) + 1 -{\ll\over 2}\right]
\en
The second equation can be solved as a function of $\ll$
\eqa
2\ll &=& 2- {w\over 2\pi m_F}+\rr\,(1-\cos\,2\th_F)
\nonumber\\
&\pm & 
\sqrt{\left(2- {w\over 2\pi m_F}+\rr\,(1-\cos\,2\th_F)\right)^2
-4\left[{\sin\,2\th_F\over 2\th_F}-{w\over 2\pi m_F}\right]}
\ena
We will not proceed further in solving those equations, but we can give a 
perturbative estimate of the shift in $\th_a$ ($w$ small):
\eq
2\dd\th_B = - {w m_F\over 8\pi m_B^2}{1\over 1-\ll}{1-\cos\,2\bar\th_F\over 
h(\bar\th_B)}
\en
\eq
2\dd\th_F = - {w \over 2\pi m_F}{1-\ll\over h(\bar\th_F)-\ll\rr\sin\,2\bar\th_F}
\en
where $h(x)=d/dx (\sin\,2x/2x)$, $\th_a = \bar \th_a + \dd\th_a$.

\setcounter{equation}{0}
\section{Conclusions}
We have shown the validity and the flexibility of the bilocal field 
method recently proposed~\cite{MC};
we have reproduced via the path-integral approach all the known results about
models as $O(N)$ and massive Gross-Neveu model in the large-$N$ limit, and
we have generalized the 't Hooft equations for $QCD_2$ and scalar $QCD_2$ to
the case of self-interacting fields.\\

Our formalism is a kind of generalization of the one used in the 
context of the $O(N)$ models~\cite{PDV}, but also applies to all cases
in which the field is a `vector' in the colour index, independently
if the vector index belongs to an internal symmetry group (such as 
$O(N)$) or to a gauge group.\\

We could also conclude that the two-dimensional gauge models with
matter in the fundamental representation are as trivial, from the point of 
view of the large-N expansion, as the $O(N)$ and $CP^{N-1}$ models, as
we just have to pay the price of introducing a bilocal colour-singlet
field instead of a local one.
This is not the case of matter in the adjoint representation, in which
one can construct an infinite number of multilocal fields that all 
contribute for large $N$.
Such a structure enlarges the difficulties but gives
a richer spectrum of states, both in the case of $QCD_2$ with 
fermionic~\cite{KT} or bosonic~\cite{DK} matter in adjoint representation.
\\
\\
{\bf Acknowledgements}\\
I would like to thank Nordita for the hospitality given
to me during the first six months of 1993.\\
I also want to thank  P. Di Vecchia for many stimulating discussions and 
for having carefully read this manuscript.


\end{document}